\documentclass[showpacs,showkeys,aps,prd,amsfonts,eqsecnum,nofootinbib,
floatfix,natbib]{revtex4-1}
\pdfoutput=1
\usepackage[T1]{fontenc}
\usepackage{amsmath,amssymb}
\usepackage{graphicx}
\usepackage{cancel}
\usepackage{float}
\usepackage{tikz}
\usetikzlibrary{decorations.pathmorphing,decorations.markings}
\tikzset{
photon/.style={decorate, decoration={snake,amplitude=2pt, segment length=5pt}, draw=black},
particle/.style={draw=black, postaction={decorate}, decoration={markings,mark=at position .5 with {\arrow[draw=black]{>}}}},
antiparticle/.style={draw=black, postaction={decorate}, decoration={markings,mark=at position .5 with {\arrow[draw=black]{>}}}},
gluon/.style={decorate, draw=black, decoration={coil,amplitude=4pt, segment length=5pt}}
goldstone/.style={draw=green,postaction={decorate},decoration={markings,mark=at position .5 with {\arrow[draw=blue]{>}}}}
}
\usepackage{feynmp-auto}
\begin{document}
\preprint{HIP-2018-15/TH, IP/BBSR/2018-10}
\title{Probing pseudo--Goldstone dark matter at the LHC}
\author{Katri Huitu}
\email{katri.huitu@helsinki.fi}
\affiliation{Department of Physics, and Helsinki Institute of Physics, P. O. Box 64, FI-00014 University of Helsinki, Finland}
\author{Niko Koivunen}
\email{niko.koivunen@helsinki.fi}
\affiliation{Department of Physics, and Helsinki Institute of Physics, P. O. Box 64, FI-00014 University of Helsinki, Finland}
\author{Oleg Lebedev}
\email{oleg.lebedev@helsinki.fi}
\affiliation{Department of Physics, and Helsinki Institute of Physics, P. O. Box 64, FI-00014 University of Helsinki, Finland}
\author{Subhadeep Mondal}
\email{subhadeep.mondal@helsinki.fi}
\affiliation{Department of Physics, and Helsinki Institute of Physics, P. O. Box 64, FI-00014 University of Helsinki, Finland}
\author{Takashi Toma}
\email{takashi.toma@gauge.scphys.kyoto-u.ac.jp}
\affiliation{Department of Physics, Kyoto University, Kyoto 606-8502, Japan}
%%%%%%%%%%%%%%%
\begin{abstract}
%%%%%%%%%%%%%%%%%%%%%%%%%%%%%%%%%%%%%%%%%%%%%%
Pseudo--Goldstone dark matter  coupled to the Standard Model  via the Higgs portal  offers an attractive 
framework for phenomenologically viable pseudo-scalar dark matter. It enjoys natural suppression of the 
direct detection rate due to the vanishing of  the relevant  (tree level) Goldstone boson vertex at zero 
momentum transfer, which makes light WIMP--like dark matter consistent with the strong current bounds.   
In this work, we explore prospects of detecting  pseudo--Goldstone dark matter at the LHC, focusing on 
the vector boson fusion (VBF) channel with missing energy. We find that, in substantial regions of 
parameter space, relatively light dark matter  ($m_\chi < 150$ GeV) can be discovered in the high 
luminosity run  as long as it is produced in decays of the Higgs--like bosons.
%%%%%%%%%%%%%%%%
\end{abstract}
%%%%%%%%%%%%%%%%%%%%%%%%
\maketitle
%%%%%%%%%%%%%%%%%%%%
\section{Introduction}
\label{sec:intro}
%%%%%%%%%%%%%%%%%%%%%
The dark matter (DM) puzzle remains one of the pressing issues in modern physics. Various particle physics 
models have been constructed which fit known properties of dark matter. Among these,  the weakly interacting 
massive particle (WIMP) paradigm remains one of the frontrunners. 

\begin{figure}[h]
\begin{center}
\includegraphics[width=5.5cm,height=3cm]{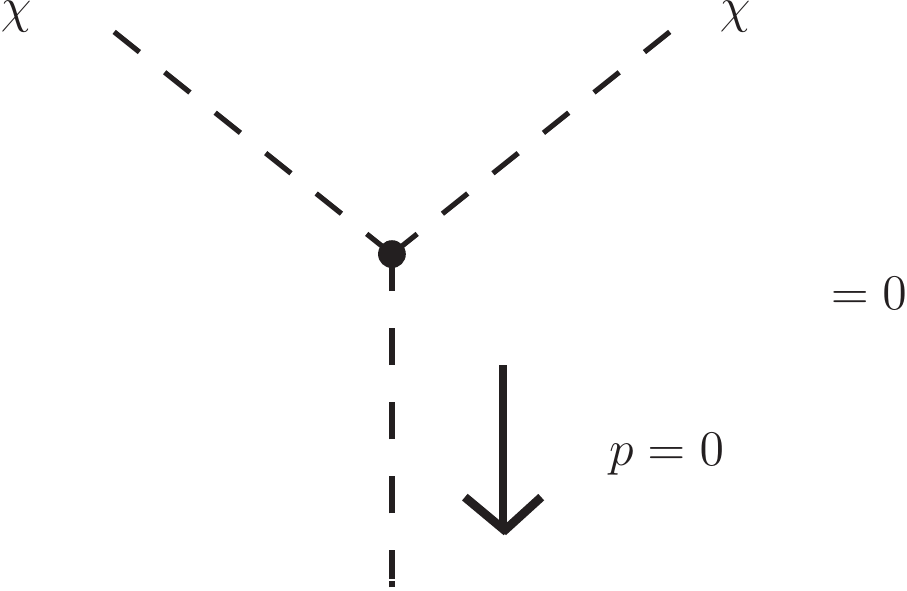}
\end{center}
\caption{The vertex involving a (pseudo-)Goldstone boson $\chi$  on-shell and a scalar vanishes at zero momentum transfer.  
This persists for a massive $\chi$ as well \cite{Gross:2017dan}.
\label{fig:mom_trans}}
\end{figure}

Recently, the electroweak--scale dark matter models have come under increasing pressure from direct detection 
experiments which have so far found null results \cite{Aprile:2018dbl,Cui:2017nnn}. These constrain the 
nucleon--dark matter interactions at (effectively) zero momentum transfer. 
An interesting option to evade such bounds is to employ the property
of Goldstone bosons that the relevant vertices vanish at zero momentum transfer (Fig.~\ref{fig:mom_trans}), 
while otherwise are unsuppressed \cite{Gross:2017dan}. 
It is important that this statement also applies to $massive$ pseudo--Goldstone bosons, which allows one to 
use this mechanism to suppress direct detection rates of WIMP--like dark matter. In this case, the nucleon--dark 
matter interaction arises at one loop level and  satisfies the XENON1T bound naturally.
%%%%%%%%%%%%%%%%%%%%%%%%%%%%%%%%%%%%%%
%%%%%%%%%%%%%%%%%%%%%%%%%%%%%%%%%%%%%%%
%%%%%%%%%%%%%%%%%%%%%%%%%%%%%%%%%%%%%%%
%%\begin{minipage}{\linewidth}
%%     \centering
%\vspace{0.6cm}      
%%\begin{minipage}{0.65\linewidth}
%\begin{figure}[H]
%\begin{center}
%\begin{tikzpicture}[thick,scale=1.0]
%\draw[dashed] (-1,1) -- node[black,above,xshift=-0.85cm,yshift=0.2cm] {$\chi$} (0,0);
%\draw[dashed] (0,0) -- node[black,above,xshift=0.8cm,yshift=0.2cm] {$\chi$} (1,1);
%\draw[dashed] (0,0) -- node[black,above,yshift=-0.3cm,xshift=1.2cm] {} (0,-1.6);
%\draw (0.5-0.25,-0.5) -- node[black,above,yshift=-0.4cm,xshift=0.7cm] {$p=0$} (0.5-0.25,-1);
%\draw (0.5-0.25,-0.5) -- node[black,above,yshift=0.3cm,xshift=3.0cm] {$=0$} (0.5-0.25,-1);
%\draw (0.375-0.25,-0.875) -- node[black,above,yshift=-0.8cm,xshift=0.9cm] {} (0.5-0.25,-1);
%\draw (0.5-0.25,-1) -- node[black,above,yshift=-0.8cm,xshift=0.9cm] {} (0.625-0.25,-0.875);
%\fill[black] (0,0) circle (0.06cm);
%\end{tikzpicture}
%\end{center}
%\caption{The vertex involving a (pseudo-)Goldstone boson $\chi$  on-shell and a scalar vanishes at zero momentum transfer.  
%This persists for a massive $\chi$ as well \cite{Gross:2017dan}. }
%\label{fig:mom_trans}
%\end{figure}
%%      \end{minipage}
%   % \hspace{0.1\linewidth}
%%       \end{minipage}
%%%%%%%%%%%%%%%%%%%%%%%%%%%%%%%%%%%%%%%

In this work, we explore the prospects of probing pseudo--Goldstone DM\footnote{Pseudo--Goldstone DM in 
a different context has been considered in 
\cite{Ibe:2009dx,Frigerio:2011in,Frigerio:2012uc,Jaeckel:2013uva,Alanne:2014kea,
Mambrini:2015nza,Carmona:2015haa,Barducci:2016fue,Ballesteros:2017xeg,Balkin:2017aep,Balkin:2018tma}.} 
at the LHC. A promising channel with a relatively low background is vector boson fusion (VBF)
Higgs production followed by its invisible decay into DM pairs \cite{Djouadi:2011aa}. The minimal model 
contains two Higgs--like scalars whose invisible decays contribute to this missing $E_T$ signature.  
Since pseudo--Goldstone bosons are naturally   light, these channels are expected to be allowed kinematically. 
The relevant constraints that restrict the efficiency of DM production are due to the current bounds on the Higgs 
invisible decay, heavy Higgs searches and indirect DM detection.  We find that despite the strong constraints, 
light DM can be probed efficiently in the high luminosity run of the LHC. 
%%%%%%%%%%%%%%%%%%%%%%
\section{Overview of the model and existing constraints}
\label{sec:model_const}
%%%%%%%%%%%%%%%%%%%%%
Consider a simple  extension of the Standard Model with a complex scalar field $S$
carrying  a global U(1) charge \cite{Gross:2017dan}. 
The most general renormalizable scalar potential invariant under global U(1) transformations $S \rightarrow e^{i\alpha} S$  is
\begin{equation}
V_0=-\frac{\mu^2_H}{2} H^\dagger H -\frac{\mu^2_S}{2} S^\ast S + \frac{\lambda_H}{2}(H^\dagger H)^2 + \lambda_{HS} H^\dagger HS^\ast S + \frac{\lambda_S}{2}(S^\ast S)^2. 
\end{equation}
The U(1) gets broken spontaneously when S acquires a vacuum expectation
value (VEV). This would result in the presence of a massless Goldstone boson. 
To avoid it, we introduce a soft breaking mass term for $S$:
\begin{equation}
V_{\rm soft}=-\frac{{\mu'_S}^{\!\!2}}{4} S^2+\rm{h.c.}
\end{equation}
with the  full scalar potential being
\begin{equation}
V=V_0+V_{\rm soft}~.
\end{equation}
In this case, a non--zero $S$ VEV generates a {\it pseudo--Goldstone} boson with mass $\mu^\prime_S$.

The parameter ${\mu'_S}^{\!\!2}$ can always be made real and positive by phase redefinition. 
The scalar potential $V$ is therefore invariant under "CP-symmetry"
\begin{equation}
S\to S^\ast ~.
\end{equation}
It is easy to show that $S$ develops a real VEV such that this symmetry is unbroken by the vacuum \cite{Gross:2017dan}. 
As a result, Im$S$ couples to other fields in pairs and is therefore stable. 

The scalar fields are parametrized as
\begin{equation}
H=\frac{1}{\sqrt{2}}\left(\begin{array}{c}
0\\
v+h
\end{array}
\right)
\end{equation}
and
\begin{equation}
S=\frac{1}{\sqrt{2}}(v_s+s+i\chi).
\end{equation}
The  stability of the pseudoscalar $\chi$ is guaranteed by 
the "CP-symmetry" and it will play the role of pseudo--Goldstone dark matter in our model.

The potential minimization conditions read
\begin{eqnarray}
&&\mu^2_H=\lambda_H v^2+\lambda_{HS} v_s^2,\\
&&\mu^2_S=\lambda_{HS} v^2+\lambda_S v_s^2-{\mu'_S}^{\!\!2}.
\end{eqnarray} 
Using these relations the parameters $\mu^2_H$ and $\mu^2_S$ can be eliminated from the scalar potential. 
The CP-even scalars $h$ and $s$ mix due to the presence of the portal coupling $\lambda_{HS}$ in the potential 
and the mass matrix in the $(h,s)$-basis is
\begin{equation}
\mathcal{M}^2=\left(\begin{array}{cc}
\lambda_H v^2     & \lambda_{HS}v v_s\\
\lambda_{HS}v v_s & \lambda_{S}v_s^2
\end{array}
\right). 
\end{equation}
It is diagonalized by the orthogonal transformation
\begin{equation}
O^T\mathcal{M}^2 O =\left(\begin{array}{cc}
m_{h_1}^2 & 0\\
0         & m_{h_2}^2
\end{array}
\right),
\end{equation}
with
\begin{equation}
O=\left(\begin{array}{cc}
\cos\theta  & \sin\theta\\
-\sin\theta & \cos\theta
\end{array}
\right),
\end{equation}
where  $\theta$ is the mixing angle, 
\begin{equation}
\tan 2\theta=\frac{2\lambda_{HS}vv_s}{\lambda_S v_s^2-\lambda_H v^2} ~.
\end{equation}
The eigenvalues of the mass--squared matrix are
\begin{equation}
m_{h_1, h_2}^2=\frac{1}{2}\left(\lambda_H v^2+\lambda_S v_s^2\mp \frac{\lambda_s v_s^2-\lambda_H v^2}{\cos 2\theta}\right).
\end{equation} 
Here $h_1$ is identified with the SM--like Higgs boson observed at the LHC.
The pseudoscalar mass  is given by the soft mass term:
\begin{equation}
m_\chi^2={\mu'_S}^{\!\!2} ~.
\end{equation}

The 6 input parameters of the scalar potential are subject to 2 experimental constraints: $v=246$ GeV and $m_{h_1}=125$ GeV, leaving 
4 free parameters which we choose as $m_{h_2}$, $\theta$, $v_s$ and $m_\chi$.
 
We note that a number of variations of the Higgs portal model with a complex scalar singlet have been considered in the 
literature \cite{McDonald:1993ex,Barger:2008jx,Gabrielli:2013hma,Gonderinger:2012rd,Grzadkowski:2018nbc}.
%%%%%%%%%%%%%%%%%%%%%%%%%%%%%%%%%%%%%%%%%%%%%%%%%%%%%%%%%%%%%%%% 
\subsection{U(1) breaking couplings as spurions}
\label{sec:u1hd_ops}
%%%%%%%%%%%%%%%%%%%%%%%%%%%%%%%%%%%%%%%%%%%%%%%%%%%%%%%%%%%%%%%% 
Here we provide a rationale for our choice of the U(1) breaking couplings following  \cite{Gross:2017dan}.
The main feature is that the odd powers of $S$ are absent and the higher even powers of $S$ are suppressed.
This can be justified by treating our set--up as the low energy limit of a more fundamental theory containing
additional states at a high energy scale. 

The simplest option would be to introduce a heavy  complex singlet
$\Phi$ which carries an even U(1) charge $q_\Phi$, while assigning $S$ an odd charge $q_S$. 
The couplings allowed by U(1) symmetry  involve even powers of $S$, e.g.
\begin{equation}
{S^{2k} \Phi^l \over \Lambda^{2k+l-4}} ~,
\end{equation}
where $k,l$ are integer and $\Lambda$ is some high energy scale.
The  (gauged) U(1) is  broken at the high scale 
by a VEV of $\Phi$ to a $Z_2$ subgroup which acts on $S$ as $S \rightarrow -S$. Define 
 \begin{equation}
 n\equiv -2 \; {q_S \over q_\Phi} ~~,~~ \epsilon \equiv \frac{\langle \Phi \rangle}{\Lambda} \,,
\end{equation}
where $\epsilon \ll 1$.
We then find 
 \begin{equation}
\mu_S^{\prime 2} \sim \langle \Phi \rangle^2 \epsilon^{n-2} ~~,~~
\lambda_{HS}^\prime \sim \lambda_S^{\prime \prime} \sim \epsilon^{n} ~~,~~ \lambda_S^\prime \sim \epsilon^{2n} \;,
\end{equation}  
where $\lambda_{HS}^\prime, \lambda_S^{\prime \prime}, \lambda_S^\prime$ are the coefficients of the 
 $|H|^2 S^2,  |S|^2 S^2, S^4$ terms in the scalar potential, respectively. Clearly, for small $\epsilon$, 
$\mu_S^{\prime 2}$ is the leading term, while the others are highly suppressed. We thus recover our effective set--up as the low energy limit.

In this example, one may require CP symmetry, in which case dark matter stability is automatic and enforced by $S \rightarrow S^*$. 
If CP is violated, dark matter is unstable and its decay is induced by the complex phases in the  higher--dimension operators. 
For  $\epsilon \ll 1$, its lifetime can be made much longer than the age of the Universe \cite{Gross:2017dan}, such that it can be 
treated as stable for all practical purposes.

We thus conclude that our assumptions can easily  be   justified by treating the U(1) breaking couplings as spurions.
%%%%%%%%%%%%%%%%%%%%%%%%%%%%%%%%%%%%%%%%%%%%%%%%%%%%%%%%%%%%%%%% 
\subsection{Cancellation of the direct DM detection amplitude}
\label{sec:dm_ddconst}
\begin{figure}[h]
\begin{center}
\includegraphics[width=3.5cm,height=5cm]{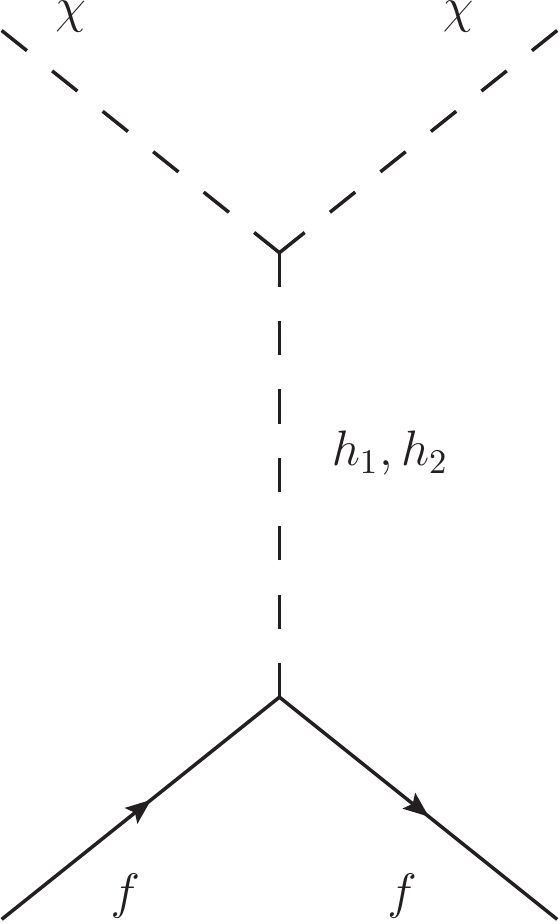}
\end{center}
\caption{DM--nucleon scattering at tree level.
\label{fig:ddcancl}}
\end{figure}

%%%%%%%%%%%%%%%%%%%%%%%%%%%%%%%%%%%%%%%%%%%%%%%%%%%%%%%%%%%%%%%%%%%%%%%%%%%%%%%%%%%%
The main feature of our dark matter framework is that the direct detection amplitude vanishes at tree level and zero momentum 
transfer \cite{Gross:2017dan}. The leading DM--nucleon scattering process is shown in Fig.~\ref{fig:ddcancl}.  Let us examine 
how the cancellation comes about in the mass eigenstate basis.
%%%%%%%%%%%%%%%%%%%%%%%%%%%%%%%%%%%%%%
%%%%%%%%%%%%%%%%%%%%%%%%%%%%%%%%%%%%%%%

The relevant interaction terms are given by
\begin{eqnarray}
&& {\mathcal L}\supset  \frac{1}{2 v_s}\chi^2 ~\biggl(  h_1\; {m^2_{h_1}}{\rm sin} \theta  -  h_2 \;{m^2_{h_2}}{\rm cos}\theta  \biggr) \;,
\label{eq:selfint1} \\
&& {\mathcal L}\supset  -\Bigl(h_1\;{\rm cos}\theta + h_2\;{\rm sin}\theta \Bigr)\sum_f \frac{m_f}{v}\bar ff \;,
\end{eqnarray}
where $f$ denotes the SM fermions.  
Thus, the tree-level direct-detection scattering amplitude is given by
\begin{eqnarray}
{\mathcal A}_{dd} (t) \propto {\rm sin}\theta ~{\rm cos}\theta\left(\frac{m^2_{h_2}}{t-m^2_{h_2}} - \frac{m^2_{h_1}}{t-m^2_{h_1}}\right)\simeq 
{\rm sin}\theta ~{\rm cos}\theta~\frac{t(m^2_{h_2}-m^2_{h_1})}{m^2_{h_1}m^2_{h_2}} \simeq 0
\label{eq:ddamp}
\end{eqnarray}
since the momentum transfer in this process is negligibly small,  $t\simeq 0$.
Thus, the contributions from $h_1$ and $h_2$ exchange   cancel each other up to 
tiny corrections of order $t/(100~{\rm GeV})^2$. This cancellation does not 
require any special relation between $m_{h_1}$ and $m_{h_2}$ and occurs  for any parameter choice.
It is of course a result of the pseudo--Goldstone nature of our dark matter. In terms of the polar coordinates,
$S=\rho e^{i \phi}$, where $\phi$ is identified with dark matter,  one finds that the $\phi \phi \rho$ vertex vanishes for
$\phi$ on shell and zero momentum of $\rho$.  This statement is specific to the explicit U(1) breaking by a mass term $S^2$
and does not hold for higher dimensional operators.
%\footnote{As detailed in \cite{Gross:2017dan}, this effective set--up can 
%be obtained from a UV theory in which U(1) is broken at high energies to $Z_2$ by a VEV of another scalar carrying an even charge. 
%This residual $Z_2$ forbids odd powers of $S$ in the potential, e.g. the linear term, while higher even--dimension operators are 
%suppressed by a high scale. } 

The cancellation is spoiled by loop corrections which generate U(1) breaking terms of dimension 4, e.g. $S^4$.
The resulting direct detection cross section is in the ballpark of $10^{-49}$ cm$^2$ \cite{Gross:2017dan}, 
which is significantly below the current bounds. A detailed analysis of these loop corrections has recently 
been performed in \cite{Azevedo:2018exj,Ishiwata:2018sdi}.
 
In our framework, the observed DM relic density  can have both thermal and non--thermal origin.  The DM annihilation cross section 
does not suffer from the above cancellation since the momentum transfer is large in this case. Thus, the correct relic abundance 
can be achieved through the usual WIMP annihilation mechanism \cite{Gross:2017dan}. In this work however, we consider a more general 
possibility that the DM production mechanism may be non--thermal, which allows for a wider range of DM masses including $m_\chi$ as 
low as 10 GeV. 
%%%%%%%%%%%%%%%%%%%%%%%%%%%%%%%%%%%%%%%%%%%%%%%%%%%%%%%%%%%%%%%%%%%%%%%%%%%%%%%%%%%%
\subsection{Invisible decay branching ratio}
\label{sec:decbr}
%%%%%%%%%%%%%%%%%%%%%%%%%%%%%%%%%%%%%%%%%%%%%%%%%%%%%%%%%%%%%%%%%%%%%%%%%%%%%%%%%%%%
In this study, we aim to probe the invisible decay of the   CP-even Higgses $h_1$ and $h_2$ 
into a pair  of DM particles $\chi$. When such decays are allowed, the VBF Higgs production 
with missing energy provides a promising channel for dark matter detection.  The $\chi$-$\chi$-$h_{1,2}$ couplings are given by
\begin{eqnarray}
&& \kappa_{\chi\chi h_1} = \frac{m^2_{h_1}}{2v_s} ~{\rm sin}\theta, \nonumber \\
&& \kappa_{\chi\chi h_2} = - \frac{m^2_{h_2}}{2v_s}~{\rm cos}\theta 
\label{eq:selfint}
\end{eqnarray} 
leading to  the invisible decay widths  
\begin{eqnarray}
&& \Gamma(h_1\to\chi\chi) = \frac{m^3_{h_1}{\rm sin}^2\theta\sqrt{1-4\frac{m^2_{\chi}}{m^2_{h_1}}}}{32\pi v_s^2}~, \nonumber \\
&& \Gamma(h_2\to\chi\chi) = \frac{m^3_{h_2}{\rm cos}^2\theta\sqrt{1-4\frac{m^2_{\chi}}{m^2_{h_2}}}}{32\pi v_s^2} ~.
\end{eqnarray}
The decay $h_1\to\chi\chi$ is quite constrained by  the LHC Higgs data,  with BR($h_1\to\chi\chi$) not exceeding 
10\% as required by the Higgs signal strength observations \cite{Khachatryan:2016vau}. On the other hand, the 
decay $h_2\to\chi\chi$ can be very efficient. It is important to include both of these DM production channels 
since either of them can dominate depending on the parameter choice.
%%%%%%%%%%%%%%%%%%%%%%%%%%%%%%%%%%%%%%%
\begin{figure}[hbt]
\begin{center}
\includegraphics[width=7.7cm,height=7.7cm]{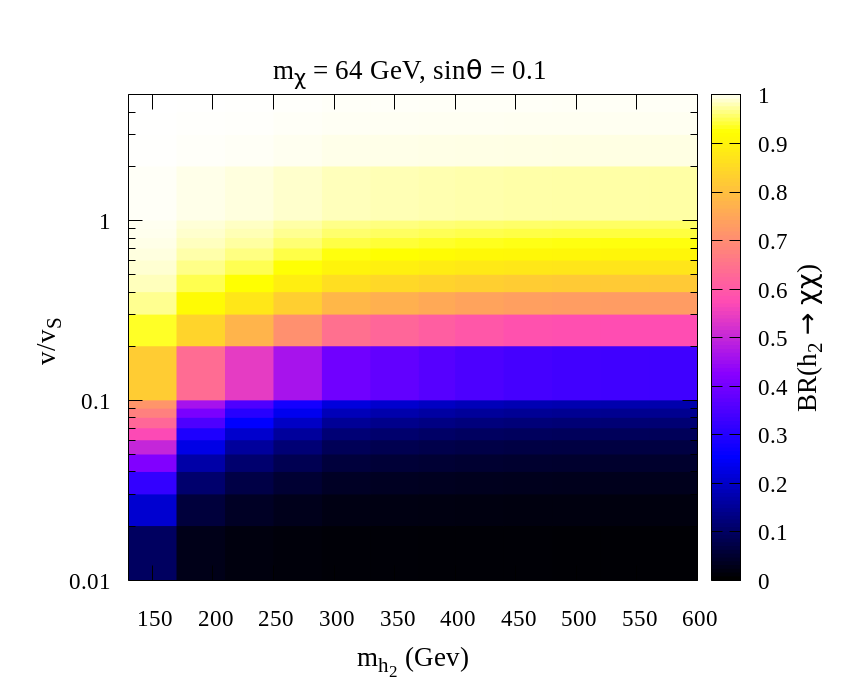}
\includegraphics[width=7.7cm,height=7.7cm]{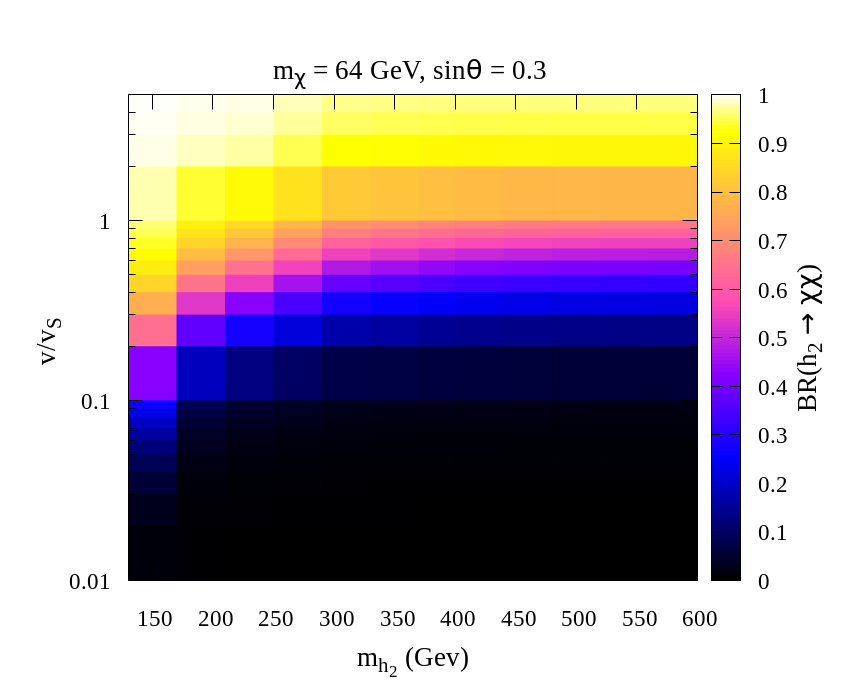}
\end{center}
\caption{Variation of BR($h_2\to\chi\chi$) in the  $(m_{h_2},v/v_s)$ plane for representative parameter values.} 
\label{fig:invbrh2}
\end{figure}
%%%%%%%%%%%%%%%%%%%%%%%%%%%%%%%%%%%%%%%

Eq.~\ref{eq:selfint} shows that the DM couplings to the CP--even scalars grow 
with $m_{h_{1,2}}$ and $1/v_s$. Given that $\sin\theta \lesssim 0.3$ as required by the  ($h_1$)  
Higgs coupling measurements \cite{Sirunyan:2018koj}, one concludes that $h_2$ couples  to DM significantly stronger  than $h_1$ does.
For $h_2$, the competing decay modes are the $\sin\theta$--suppressed  $h_2\to VV^{(*)}$ and, for larger 
$m_{h_{2}}$, $h_2\to h_1 h_1$ and $h_2\to t\bar t$. The variation of  BR($h_2\to\chi\chi$) for typical parameter choices is shown in Fig.~\ref{fig:invbrh2}.
We keep $m_{h_2}$ below 600 GeV to have a substantial $h_2$-- production cross section and choose $m_\chi=64$ GeV to evade the 
BR($h_1\to\chi\chi$)  constraint. We see that for $v/v_s >0.1$, the invisible decay mode is significant and often dominant.
%%%%%%%%%%%%%%%%%%%%%%%%%%%%%%%%%%%%%%%%
\subsection{ Constraints}
\label{sec:ind_detn}
%%%%%%%%%%%%%%%%%%%%%%%%%%%%%%%%%%%%%%%%
The model  parameters  $m_{h_2}$, $m_{\chi}$, $v_s$ and $\theta$ are constrained  by various experiments.
Perturbative unitarity considerations exclude small values of $v_s \ll v$, which for fixed scalar masses imply
large quartic couplings. The  mixing angle $\theta$ and dark matter mass $m_{\chi}$ are constrained by the LHC Higgs coupling data.  
In addition, depending on the choice of $\theta$, the heavy CP-even Higgs boson mass $m_{h_2}$ is subject to the LHC direct search  
bound.\footnote{A summary of analogous constraints in the real scalar extension of the SM can be found in \cite{Falkowski:2015iwa}.} 
While the direct DM detection constraint is weak and superseded by that from perturbative unitarity in the relevant parameter range, 
the indirect DM detection constraint from the Fermi satellite is significant for relatively light DM.
Finally, when $\chi$ is assumed to have been produced thermally, there is a PLANCK constraint on the DM annihilation cross section, 
which requires a substantial DM--scalar couplings away from the resonance regions. The dark matter computations were performed
using micrOMEGAs \cite{Belanger:2013oya}.

Below we delineate parameter space consistent with all of the constraints, keeping the spectrum at the electroweak scale and the mixing angle below 0.3. 
%%%%%%%%%%%%%%%%%%%%%%%%%%%%%%%%%%%%%%%%%%%%%%%%%
\subsubsection*{Constraints from unitarity,  invisible Higgs decay and dark matter detection experiments}
\label{sec:const_scaldm}
\begin{figure}[t]
\begin{center}
\includegraphics[scale=0.64]{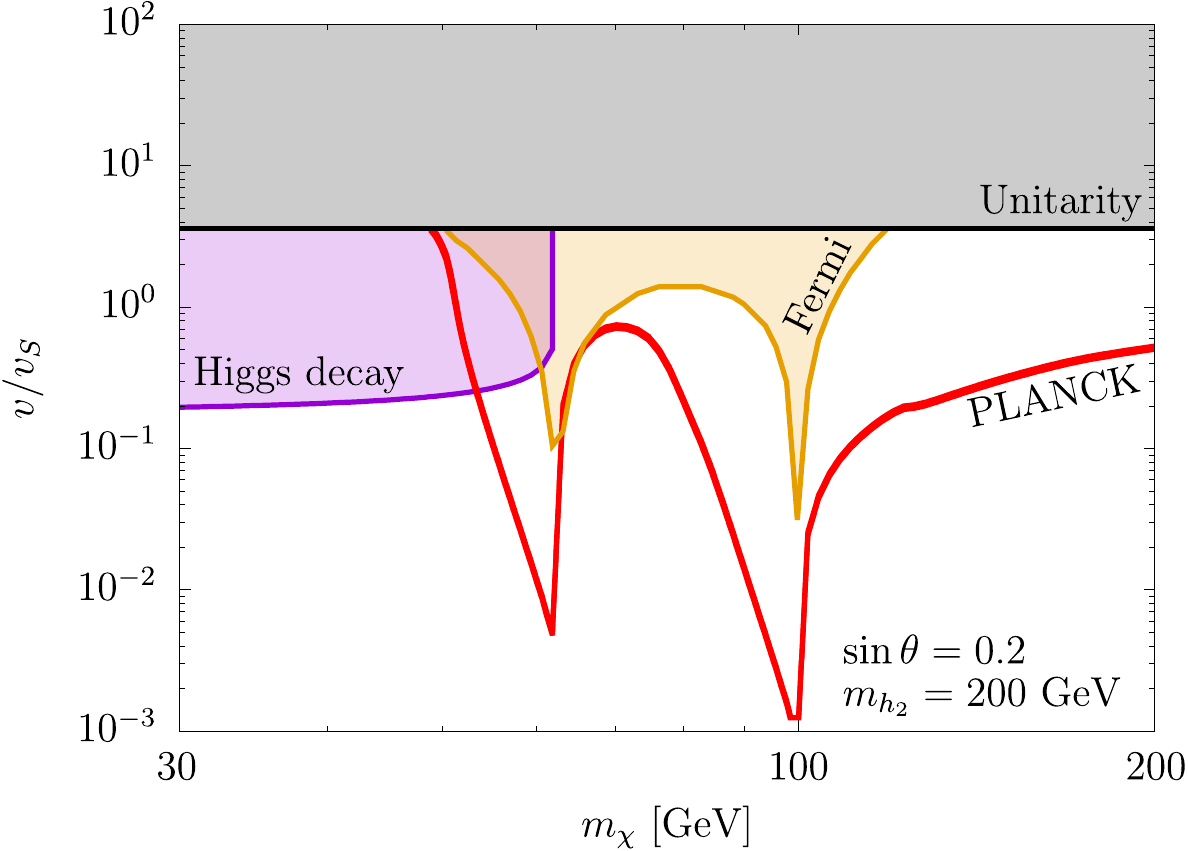} 
\includegraphics[scale=0.64]{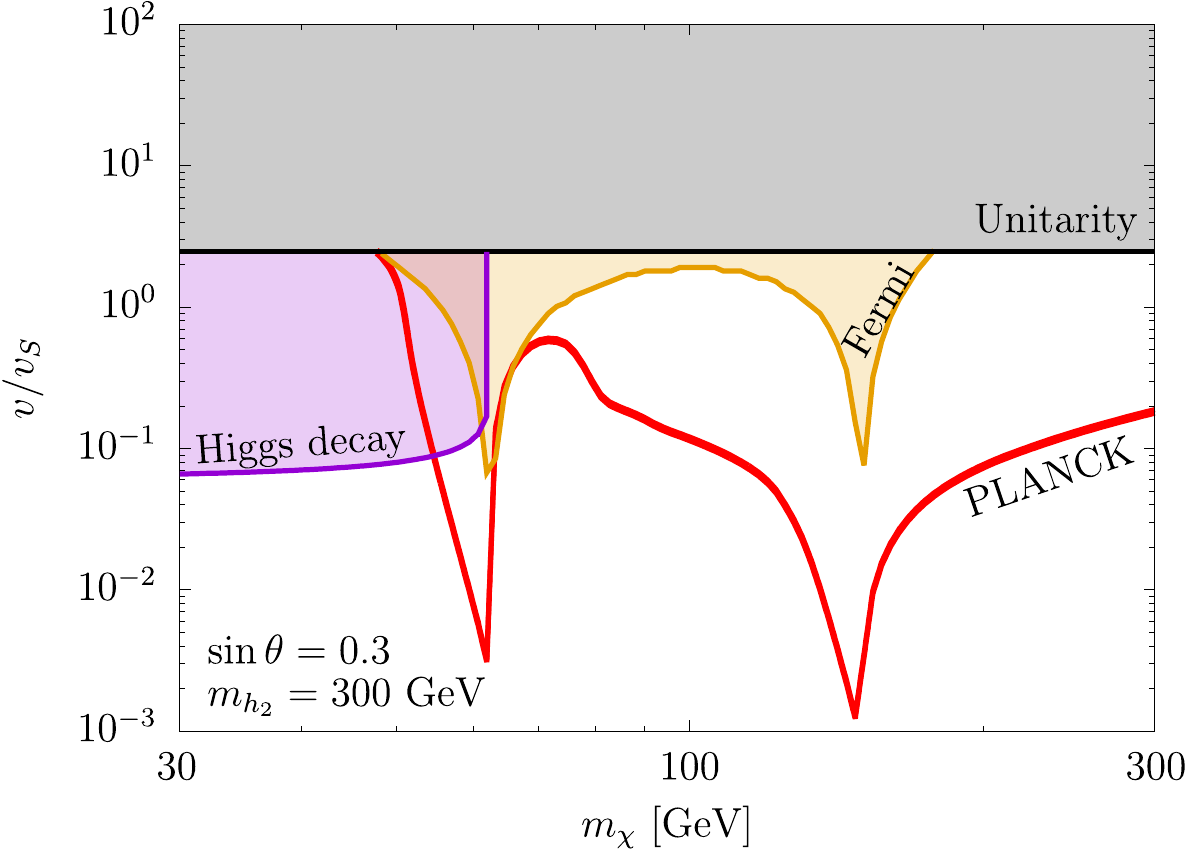}
\caption{ ($m_\chi$, $v/v_s$) parameter space.   The colored regions
are excluded by   perturbative unitarity (grey), the Higgs invisible decay
(purple)  and gamma--ray observations (orange). The red band gives the correct thermal DM relic abundance according to PLANCK. } 
\label{fig:num}
\end{center}
\end{figure}

%%%%%%%%%%%%%%%%%%%%%%%%%%%%%%%%%%%%%%%%%%%%%%%%%
Figure~\ref{fig:num} shows the results of our numerical analysis at fixed representative values of   $\sin\theta$ and $m_{h_2}$. 
The grey, purple and orange regions are excluded by the perturbative unitarity constraint $\lambda_S<8\pi/3$~\cite{Chen:2014ask}, 
the Higgs invisible decay bound~\cite{Khachatryan:2016vau} and the gamma--ray observations from dwarf spheroidal galaxies (dSphs)
with 6 years and 15 dSphs data by the Fermi--LAT Collaboration~\cite{Ackermann:2015zua,Fermi-LAT:2016uux}, respectively.

To constrain the  invisible  $h_1$ Higgs decay, we use the Higgs signal strength value $\mu = 1.09^{+0.11}_{-0.10}$  
obtained with 7 + 8 TeV LHC data 
\cite{Khachatryan:2016vau}.
In our model, the effective $\mu$ is given by
\begin{equation}
\mu = \cos^2\theta ~ (1 - {\rm BR_{inv}}) \;,
\end{equation} 
where $ {\rm BR_{inv}}$ is the $h_1$ invisible decay branching ratio. 
The bound on ${\rm BR_{inv}} $ is $\theta$--dependent:  at $\sin\theta=0.1$, it gives ${\rm BR_{inv}} <10\%$  at $2\sigma$, while for
$\sin\theta=0.3$ it strengthens to ${\rm BR_{inv}} <2\%$.
 The direct 13 TeV bound BR$_{\rm inv} < 26\%$ \cite{ATLAS:2018doi,Sirunyan:2018koj} is weaker,
while the 13 TeV constraints on the Higgs signal strength  $1.13^{+0.09}_{-0.08}$ (ATLAS) \cite{ATLAS:2018doi} and $1.17\pm 0.10$
(CMS) \cite{Sirunyan:2018koj} are only consistent with the SM at 2$\sigma$ level. It is therefore reasonable to  use a  conservative 
bound quoted above.
%%%%%%%%%%%%%%%%%%%%%%%%%%%%%%%%%%%%%%%%%%%%%%%%%
%%%%%%%%%%%%%%%%%%%%%%%%%%%%%%%%%%%%%%%%%%%%%%%%%

For the gamma--ray constraint, a Navarro--Frenk--White (NFW) dark matter profile is assumed~\cite{Navarro:1996gj}. The red
band shows the allowed parameter range for thermal dark matter, whose abundance lies within   a $3\sigma$
interval of $\Omega_{\chi}h^2=0.1197\pm0.0022$ as reported by the PLANCK Collaboration~\cite{Ade:2015xua}.
One observes the presence of the characteristic  dips associated with the  resonant DM annihilation through  
$h_1$ and $h_2$. The DM couplings are allowed to be very small around these dips. The low mass end of the red 
bands,  $m_{\chi}<m_{h_1}/2$, is excluded by the invisible Higgs decay constraint, while the Fermi bound does not 
significantly affect the allowed parameter space for thermal DM. The direct DM detection constraint is  loose and 
superseded by  that from   perturbative unitarity (grey area) \cite{Gross:2017dan}.
  
The uncolored regions are allowed by all of the constraints as long as the DM production mechanism is non--thermal. 
In particular, $m_\chi = {\cal O}(10)$ GeV is consistent with  the Higgs invisible decay bound for small DM--Higgs couplings.
%%%%%%%%%%%%%%%%%%%%%%%%%%%%%%%%%%%%%%%%%%%%%%%% 
\subsubsection*{Constraints from the LHC direct search}
\label{sec:const_scaldir}
%%%%%%%%%%%%%%%%%%%%%%%%%%%%%%%%%%%%%%%%%%%%%%%%%
Further constraints are imposed by the direct LHC search for Higgs--like states in various channels.
We have taken into account the following 13 TeV LHC results:
\begin{itemize}
\item The $h_2\to \gamma\gamma$ searches by  ATLAS \cite{Aaboud:2017yyg} and CMS \cite{Sirunyan:2018wnk}. The ATLAS search 
probes the heavy Higgs masses above 200 GeV, whereas the CMS lower bound is  500 GeV.

\item Searches for $h_2\to WW$ and $h_2\to ZZ$ in ATLAS \cite{Aaboud:2017gsl,Aaboud:2018bun,Aaboud:2017rel} and 
CMS \cite{Sirunyan:2018qlb}. The CMS  $h_2\to ZZ$ search \cite{Sirunyan:2018qlb} probes the lowest mass range: 
$m_{h_2}\geq 130$GeV. The ATLAS $h_2\to WW$ and $h_2\to ZZ$ searches are sensitive to masses  above  200 GeV. 
The combined  ATLAS limits from   $h_2\to WW$ and $h_2\to ZZ$ searches impose  a limit on $m_{h_2}\geq 300$ GeV. 

\item The ATLAS  searches for $h_2\to h_1 h_1$ in $4b$ \cite{Aaboud:2018knk}, $2\gamma$ $2b$ \cite{Aaboud:2018ftw} and 
$2W$ $2\gamma$ final states, and the CMS searches in $2\gamma$ $2b$ \cite{Sirunyan:2018iwt} and $4b$ \cite{Sirunyan:2018zkk} 
final states. The ATLAS searches and the CMS $4b$  search  probe the heavy Higgs masses   above 260 GeV and the CMS $2\gamma$ $2b$ 
search probes those above 250 GeV.

\item The ATLAS search for the invisible $h_2$ decay in the VBF production channel \cite{Aaboud:2018sfi}. 
This probes the heavy Higgs mass range starting from 100 GeV.
\end{itemize}
  
We are interested in the low mass range   150--300 GeV as this leads to  the strongest signal. 
The above searches set  a bound on the production cross section of the final state in question. 
In the Narrow Width Approximation (NWA), it is given  by:
\begin{equation}\label{production cross section}
\sigma_{\rm prod}= \sigma(pp\to h_2)\times \textrm{BR}(h_2\to\textrm{SM}).
\end{equation}  
The $h_2$ production cross section is proportional to $\sin^2\theta$,
\begin{equation}\label{h2 production cross section}
\sigma(pp\to h_2)=\sin^2\theta \times \sigma^{\rm SM}_{pp\to h}(m_h=m_{h_2}),
\end{equation}
where  $\sigma^{\rm SM}_{pp\to h}$ is the production cross section of the SM Higgs boson. The branching ratio for the  $h_2$ decay to a given SM final state is 
\begin{equation}\label{h2 br to SM final states}
\textrm{BR}(h_2\to\textrm{SM})=\frac{\sin^2\theta ~ \Gamma_{h\to \textrm{SM}}(m_h=m_{h_2})}{\sin^2\theta ~ \Gamma_{h}^{\rm tot}(m_h=m_{h_2})+\Gamma(h_2\to\chi\chi)+\Gamma(h_2\to h_1 h_1)},
\end{equation}
where $\Gamma_{h\to \textrm{SM}}$ is the  SM Higgs decay rate to the final state  and $\Gamma_{h}^{\rm tot}$ is the SM Higgs total decay width.  

\begin{figure}[t]
\begin{center}
\includegraphics[scale=0.67]{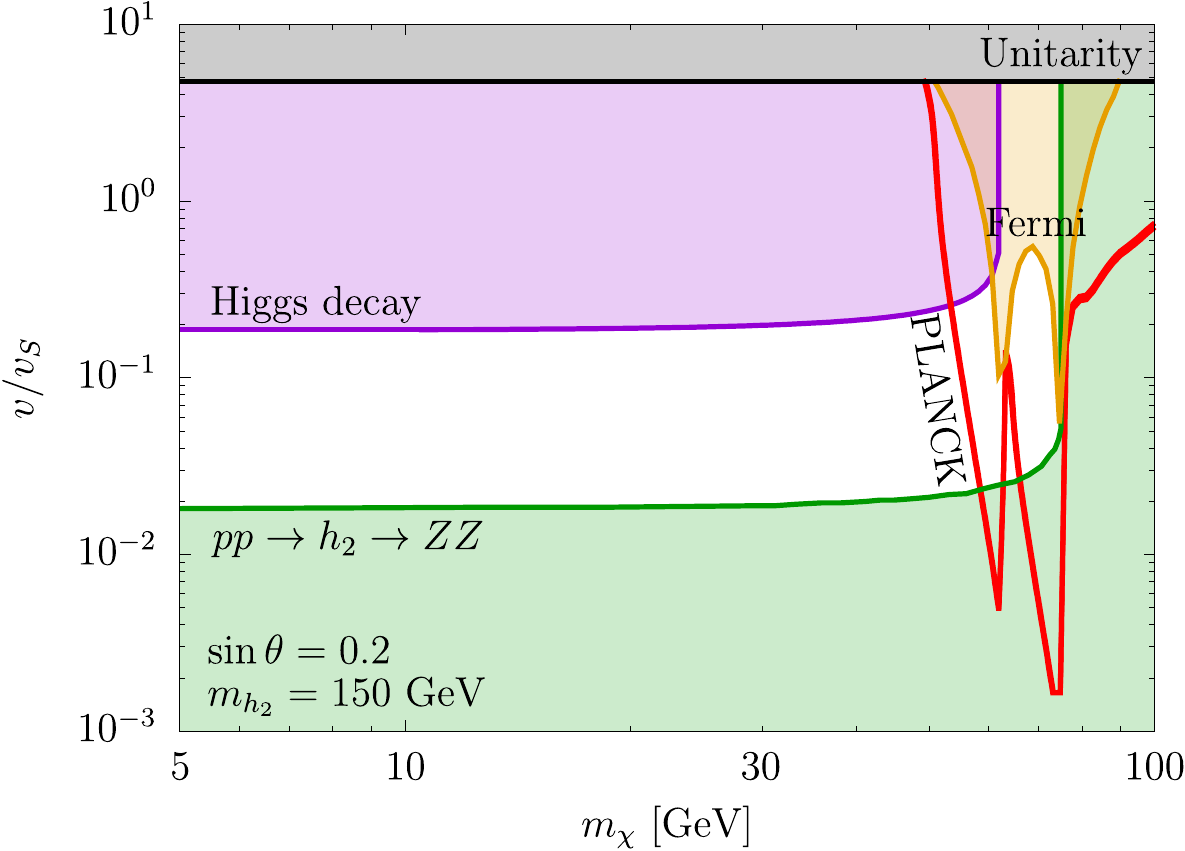}
\includegraphics[scale=0.67]{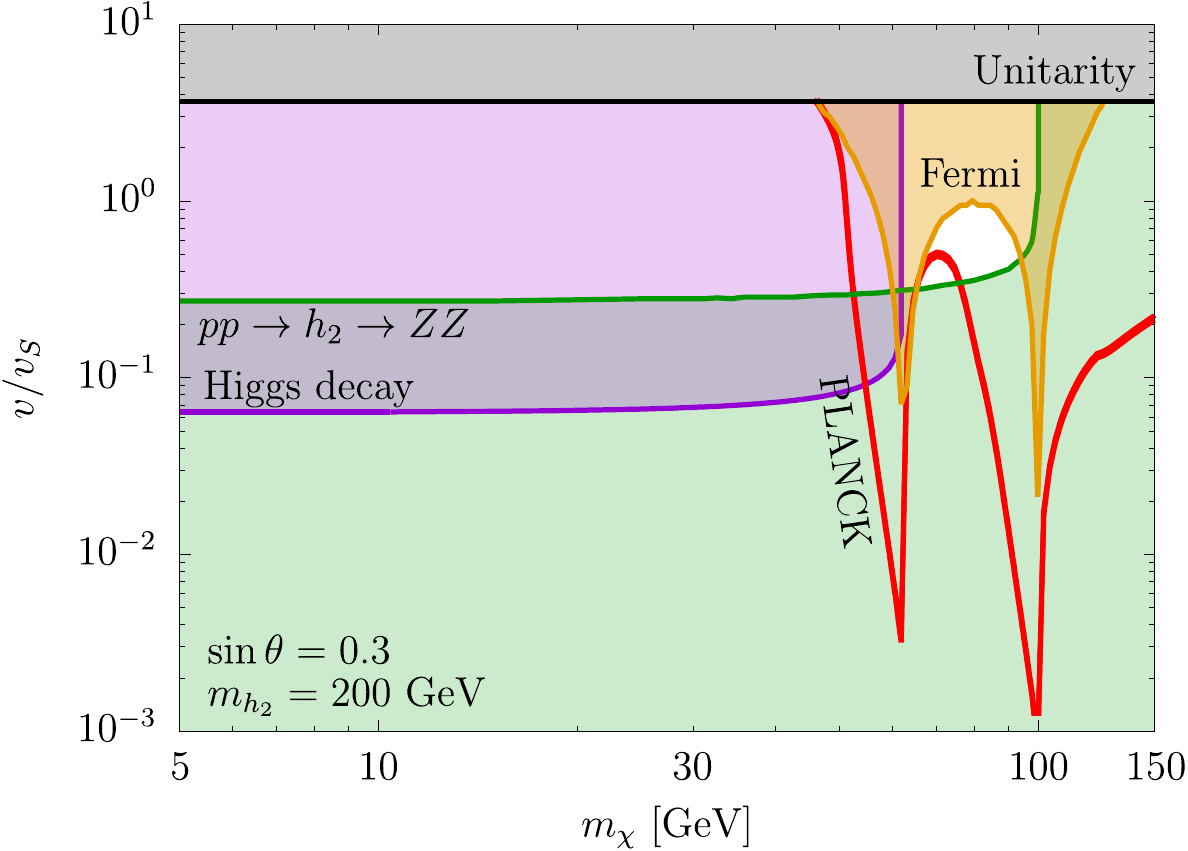}
\caption{LHC  $h_2$--search  constraints (green) in the  ($m_\chi$, $v/v_s$) plane. The other constraints are as in Fig.~\ref{fig:num}.} 
\label{fig:const2}
\end{center}
\end{figure}

It is clear  that $\sigma_{\rm prod}$ receives two suppression factors: $\sin^2\theta$ and the presence of non--standard decay channels.
For the  mixing angle values $\sin\theta=0.1$, $0.2$ and $0.3$,  the cross section  $\sigma_{\rm prod}$ is below the limit
for the $\gamma\gamma$, $WW$ and $h_1 h_1$ final states.  The $ZZ$   searches in both CMS \cite{Sirunyan:2018qlb} 
and ATLAS \cite{Aaboud:2017rel} and the combined $WW$, $ZZ$ search in ATLAS \cite{Aaboud:2018bun}  are 
more sensitive and   for $\sin\theta=0.2,0.3$   impose a non--trivial constraint shown in Fig.~\ref{fig:const2}.
%%%%%%%%%%%%%%%%%%%%%%%%%%%%%%%%%%%%%%%%%%%%%%%%%
%%%%%%%%%%%%%%%%%%%%%%%%%%%%%%%%%%%%%%%%%%%%%%%%%

Below the $h_2\to h_1 h_1$ kinematic threshold, the necessary suppression is provided by the invisible decay channel $h_2 \to \chi \chi$ as long as 
$m_{h_2}>2m_\chi$. While the $\sin\theta=0.1$ case is safe regardless  of the efficiency of the invisible decay, at $\sin\theta=0.2,0.3$ 
a large BR$_{\rm inv}(h_2)$ is required in order for a light $h_2$ to be consistent with the LHC data. This excludes low values of $v/v_s$  at which 
the $\chi$--$\chi$--$h_2$ coupling  is suppressed, as shown in Fig.~\ref{fig:const2}. 
The constraint is rather sensitive to the exact value of $m_{h_2}$: while the observed limit for $m_{h_2}=150$ GeV is lower than  the corresponding 
SM value by a factor of 31, the analogous factor for $m_{h_2}=200$ GeV is only 22. That  means, in the latter case the $\sin^2\theta=(0.2)^2 $ 
suppression is enough to satisfy the bound at any $v_s$, while in the former case an efficient invisible decay channel is necessary. 

It is interesting that at higher $m_{h_2} \sim 300$ GeV, the above constraint disappears due to the additional  decay channel $h_2\to h_1 h_1$. 
The corresponding coupling
\begin{equation}
C_{h_1 h_1 h_2} = - \frac{2m_{h_1}^2 + m_{h_2}^2}{4vv_s}  \sin 2\theta ~ (v_s{\rm cos}\theta + v{\rm sin}\theta), 
\end{equation}
is quite large and remains unsuppressed  at large $v_s$ making the standard Higgs decay channels much less efficient.
Thus, both panels of Fig.~\ref{fig:num} are consistent with the $h_2$--constraint. 
%%%%%%%%%%%%%%%%%%%%%%%%%%%%%%%%%%
\section{Collider Analysis}
\label{sec:collan}
\begin{figure}[hbt]
\begin{center}
\includegraphics[width=5cm,height=5.0cm]{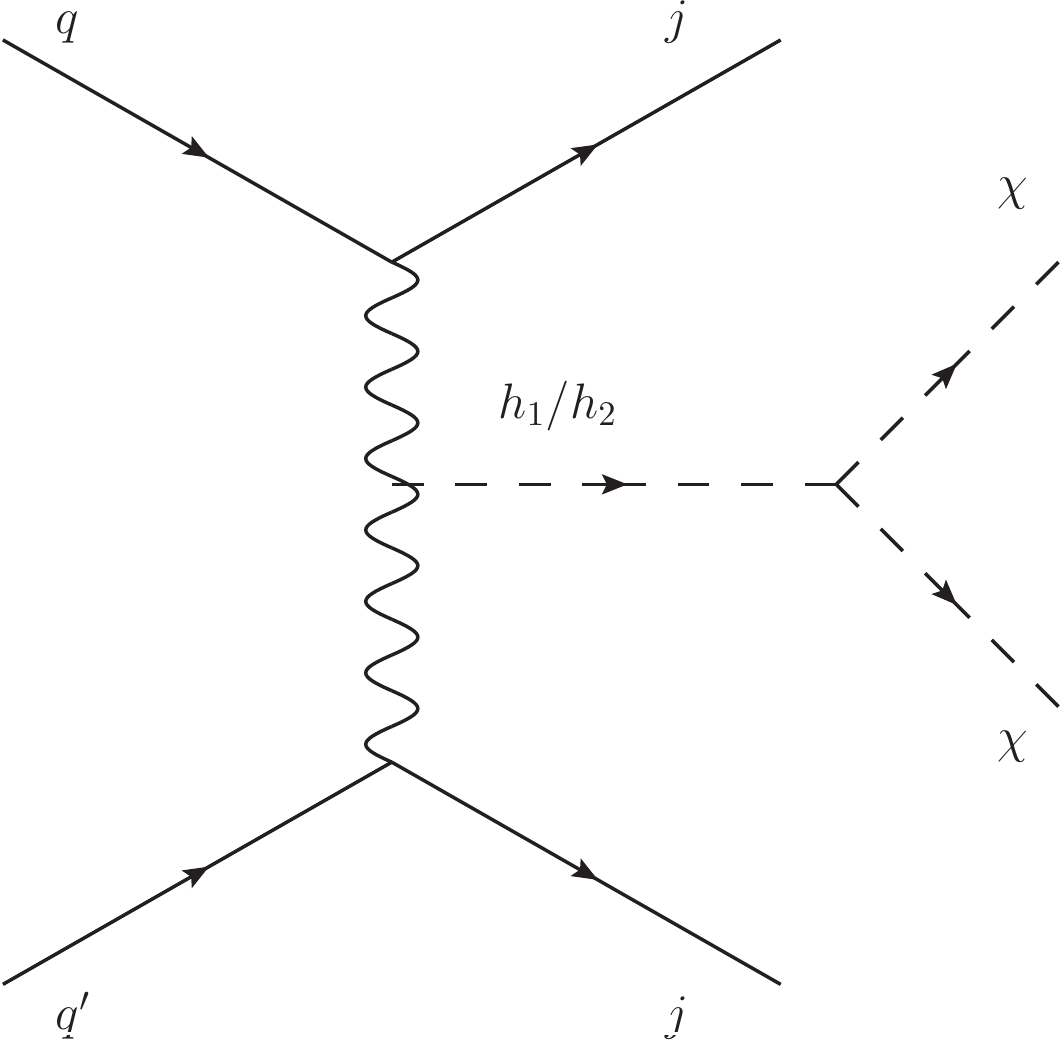}
\end{center}
\caption{DM pair production via vector boson fusion (VBF) at the LHC.
\label{fig:prodvbf}}
\end{figure}

%%%%%%%%%%%%%%%%%%%%%%%%%%%%%%%%%%
We aim to probe our DM  model via the missing energy signature in the VBF Higgs production channel (Fig.~\ref{fig:prodvbf}).
There are of course other options as well. For example, Ref. \cite{Kim:2015hda} has explored  gluon fusion with the initial 
or final state radiation accompanied by missing energy, within a somewhat different Higgs portal DM model. The problematic 
aspect in this case is the large QCD background. For the VBF mode, the background is lower since the jets produced in this 
process are forward and easier to tag on. Thus it appears to be  a promising channel \footnote{In order to ascertain 
the fact that a VBF production channel is more sensitive compared to a gluon fusion production resulting in a monojet signal,  
we made an estimate with reference to the results in \cite{Kim:2015hda}. We found that with similar parameter choices, the 
significance factor in VBF production channel can be larger by a factor $\sim$ 1.6.}.
%%%%%%%%%%%%%%%%%%%%%%%%%%%%%%%%%%
%%%%%%%%%%%%%%%%%%%%%%%%%%%%%%%%%%
 
We find that the process is efficient only when the DM pair is produced by on--shell  $ h_1$ and/or $h_2$ decays. 
Thus we will focus on the region  $m_{h_2}> 2 m_\chi$. In addition, due to the $\sin^2 \theta$ suppression, the 
heavy Higgs production cross section is significant only for the electroweak range masses. Although the $h_1$ 
production is unsuppressed by the mixing angle, the kinematic reach of $h_1 \to \chi\chi$ is smaller than that 
for the $h_2$ decay and, furthermore,  its branching ratio is already strongly constrained. As we detail below,  
both $h_1$ and $h_2$ can give a dominant contribution to DM pair production, depending on the parameter region.

%%%%%%%%%%%%%%%%%%%%%%%%%%%%%%%%%%
\subsection{Simulation details}
\label{sec:simuln}
%%%%%%%%%%%%%%%%%%%%%%%%%%%%%%%%%%
The ATLAS and CMS Collaboration have explored the possibility of detecting invisible decay of the Higgs boson 
in the VBF channel and have optimised the corresponding cuts  \cite{Sirunyan:2018owy,Aaboud:2018sfi}. The definitions 
of the signal region and  kinematic cuts  are very similar in both of the analyses. In this work, we have adopted the CMS analysis 
of the VBF channel with $\sqrt{s}=13$ TeV data \cite{Sirunyan:2018owy}. 
The VBF signal events are required to produce at least two jets with transverse momentum  $p_T > 80 (40)$ GeV  for the leading 
(subleading) jet and rapidity  $|\eta| < 4.7$. Further, at least one of the two leading jets must have 
$|\eta| > 3$. Events are also required to have a large transverse missing energy, $\cancel{E_T} > 250$ GeV.

The dominant SM background contribution to the signal region arises from the $W$ + jets and $Z$ + jets channels, with 
$W$ and $Z$ decaying leptonically into $\ell\nu_{\ell}$ and $\nu_{\ell}\bar\nu_{\ell}$, respectively. 
The $W$ + jets channel contributes due to a non-zero lepton misidentification probability combined with a large  production
cross section. The  next largest contributions are due to the  top production (single and pair) and gauge boson 
pair production. Contributions from the top production channels are suppressed by the small lepton misidentification probability, whereas the
gauge boson pair production channels suffer from smaller cross-sections. QCD jets and $\gamma$ + jets are the other two potentially large contributors 
to the background. Although they do not have any direct source of missing energy in the final state, mismeasurements  
of the momenta of the final state particles, especially jets, can result in a non-zero $\cancel{E_T}$.

In order to suppress the QCD background contribution, all the jets in the final state with $p_T > 30$ GeV and $|\eta| < 4.7$ 
are required to be separated from the missing momentum direction $\vec{\cancel{E_T}}$ by an azimuthal angle 
$\Delta\phi(\vec{\cancel{E_T}},\vec{p_T}^{\rm jet}) > 0.5$ radians. The two leading jets in the VBF channel 
are expected to be widely separated by pseudorapidity $\eta$ and almost back-to-back. This kinematic feature 
is exploited by requiring $\eta_{j1} \eta_{j2} < 0$ and $|\Delta\eta_{jj}| > 4$. These jets are also likely 
to have a large invariant mass, $m_{jj} > 1300$ GeV, and a small azimuthal separation, $|\Delta\phi_{jj}| < 1.5$. 
Such cuts are particularly effective in reducing the SM background contributions arising from the $V$+jets channel, where
$V=W, Z$. The $W$+jets background is further reduced by implementing a lepton veto. Events are rejected if they 
contain a muon (electron) with $p_T > 10$ GeV and $|\eta| < 2.4 (2.5)$ or a $\tau$-lepton decaying hadronically 
($\tau_h$) with $p_T > 10$ GeV and $|\eta| < 2.3$. Any event with a photon in the final state with $p_T > 15$ GeV 
and $|\eta| < 2.5$ is also rejected to suppress the $\gamma$+jets background. A $b$-jet veto is implemented in order 
to reduce the background from the $t\bar t$ and single top production channels. The $b$-jet candidates are required to have 
$p_T > 20$ GeV and $|\eta| < 2.4$. The combined secondary vertex algorithm \cite{Chatrchyan:2012jua,Sirunyan:2017ezt} 
adopted by the CMS Collaboration yields  roughly a 60\% efficiency of tagging the $b$-jets. A small probability (1\%) of misidentifying a light-flavor 
jet as a $b$-jet is also taken into account. We have summarised the selection criteria for the final state in Table~\ref{tab:def_sr}.
%%%%%%%%%%%%%%%%%%%%%%%%%%%%%%%%
\begin{table}[h!]
\begin{center}
\begin{tabular}{||c||c||}
\hline
Observables & Requirements \\
\hline\hline
Leading (trailing) jet & $p_T > 80 (40)$ GeV, $|\eta| < 4.7$   \\
$\cancel{E_T}$  & $> 250$ GeV   \\
$\Delta\phi(\vec{\cancel{E_T}},\vec{p_T}^{\rm jet})$  &  $> 0.5$   \\
$|\Delta\phi_{jj}|$  &  $< 1.5$   \\
$\eta_{j1}.\eta_{j2}$  &  $< 0$   \\
$|\Delta\eta_{jj}|$  &  $> 4$   \\
$|m_{jj}|$  &  $> 1300$ GeV   \\
Leptons  &  $N_{\mu ,e}=0$, $p_T > 10$ GeV, $|\eta| < 2.4(2.5)$   \\
$\tau$ leptons  &  $N_{\tau_h}=0$, $p_T > 10$ GeV, $|\eta| < 2.3$   \\
Photons  &  $N_{\gamma}=0$, $p_T > 15$ GeV, $|\eta| < 2.5$   \\
$b$-jets  &  $N_{b}=0$, $p_T > 20$ GeV, $|\eta| < 2.4$  \\
\hline\hline
\end{tabular}
\caption{Final state selection cuts for the VBF analysis  \cite{Sirunyan:2018owy}.}
\label{tab:def_sr}
\end{center}
\end{table}
%%%%%%%%%%%%%%%%%%%%%%%%%%%%%%%%

The model has been implemented in MadGraph5 \cite{Alwall:2011uj,Alwall:2014hca} through FeynRules \cite{Christensen:2008py,
Alloul:2013bka,Degrande:2011ua}. 
Event generation at the parton level is performed by MadGraph. We have used the parton distribution function
NNPDF \cite{Ball:2012cx,Ball:2014uwa} for our computation. 
These events are subsequently passed to Pythia8 \cite{Sjostrand:2014zea} for showering and 
hadronisation. Jet formation is done using the anti-kT algorithm \cite{Cacciari:2008gp} by FastJet \cite{Cacciari:2011ma} and detector simulation 
is performed using Delphes3 \cite{deFavereau:2013fsa,Selvaggi:2014mya,Mertens:2015kba}. 
Finally, the events are analysed by CheckMATE \cite{Drees:2013wra,Dercks:2016npn}.   
%%%%%%%%%%%%%%%%%%%%%%%%%%%%%%%%%%
\subsection{Results and discussion}
\label{sec:res}
%%%%%%%%%%%%%%%%%%%%%%%%%%%%%%%%%%
In this section, we discuss the results of our collider simulation for two choices of the heavy CP-even Higgs mass, 
$m_{h_2}=150$ GeV and 300 GeV. Since the $h_2$ production  is suppressed by 
small ${\rm sin}^2 \theta$, we restrict the $h_2$ mass to the electroweak range.
  
In order to estimate the statistical significance factor ${\mathcal S}$, we use 
\begin{equation}
{\mathcal S}=\frac{S}{\sqrt{S+B+\sigma_B^2}}, 
\end{equation}
where $S$, $B$ and $\sigma_B$ represent the number of signal events, SM background events and the uncertainty in the 
background measurement. The CMS Collaboration quotes $B\pm\sigma_B = 1779\pm 96$ at $35.9~{\rm fb}^{-1}$ integrated 
luminosity \cite{Sirunyan:2018owy}. To estimate ${\mathcal S}$ at high integrated luminosity, we have scaled $B$ 
accordingly and taken two choices of $\sigma_B$ to reflect our lack of knowledge of how the realistic uncertainty 
may evolve. The ``best case scenario''  corresponds to a negligible background uncertainty, $\sigma_B \ll \sqrt{B}$, 
and the second choice is $\sigma_B =\sqrt{B} $ which effectively reduces the significance by $\sqrt{2}$. We expect 
that the resulting significance gives an idea of the realistic signal strength and the detection prospects. 

Below we consider  two representative values of $m_{h_2}$ and the resulting signals.
%%%%%%%%%%%%%%%%%%%%%%%%%%%%%%%%%%
\begin{figure}[h!]
\begin{center}
\includegraphics[width=7cm,height=7.0cm]{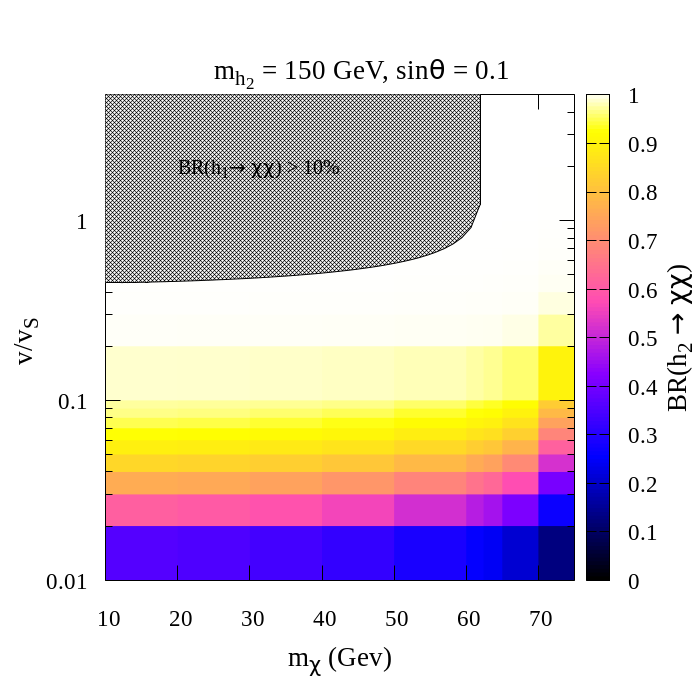}
\includegraphics[width=7cm,height=7.0cm]{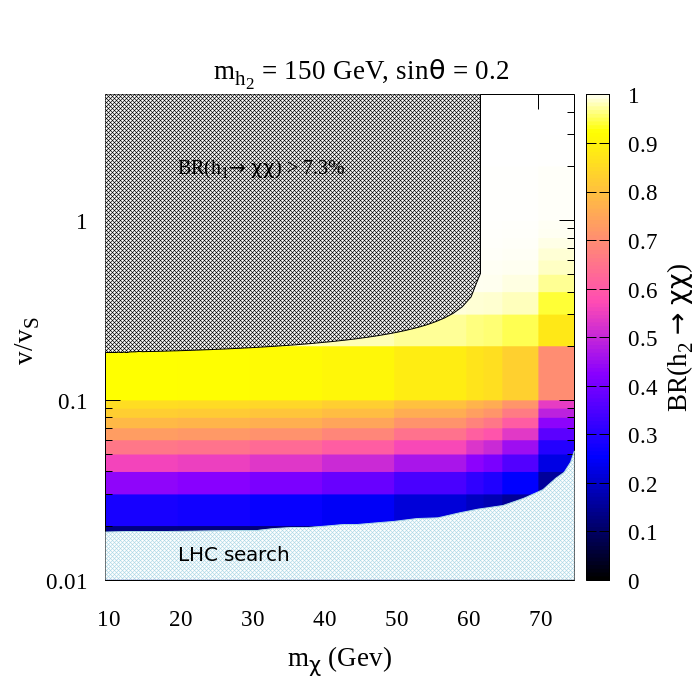}
\end{center}
\caption{BR($h_2\to\chi\chi$) and the LHC constraints for  $m_{h_2}$=150 GeV.
\label{fig:brmh150}}
\end{figure}
%%%%%%%%%%%%%%%%%%%%%%%%%%%%%%%%%%
%%%%%%%%%%%%%%%%%%%%%%%%%%%%%%%%%%
\subsubsection*{ a. $m_{h_2}=150~\mathrm{GeV}$}
\label{sec:resmh150}
%%%%%%%%%%%%%%%%%%%%%%%%%%%%%%%%%%
\noindent
We fix  ${\rm sin}\theta$ to be  0.1 and 0.2  since for  larger ${\rm sin}\theta\ge 0.3$  a 150 GeV CP-even 
Higgs is nearly excluded by the LHC data \cite{Aaboud:2017rel,Aaboud:2017gsl}. We vary the ratio ${v}/{v_s}$ 
within the range [0.01 - 5.0], where the upper limit is set by  the perturbative unitarity constraint \cite{Chen:2014ask}. The DM mass $m_{\chi}$ 
is bounded from above by ${m_{h_2}}/{2}$ as required by the on--shell decay of $h_2$ into a DM pair.

In Fig.~\ref{fig:brmh150},  we  show the variation of 
BR($h_2\to\chi\chi$) in the plane (${v}/{v_s}$, $m_{\chi}$) for two choices of ${\rm sin}\theta$. 
As Eq.~\ref{eq:selfint} suggests, the branching ratio  increases with smaller ${\rm sin}\theta$ and  ${v_s}$. 
The upper left corner in both  figures is excluded  by the invisible decay of the 125 GeV 
Higgs boson, while the low values of $v/v_s$ in the right panel are ruled out by the $h_2$ searches. We see that  
large values of  BR$_{\rm inv}(h_2)$ close to one are consistent with the LHC constraints.

\begin{figure}[h!]
\begin{center}
\includegraphics[width=7cm,height=7.0cm]{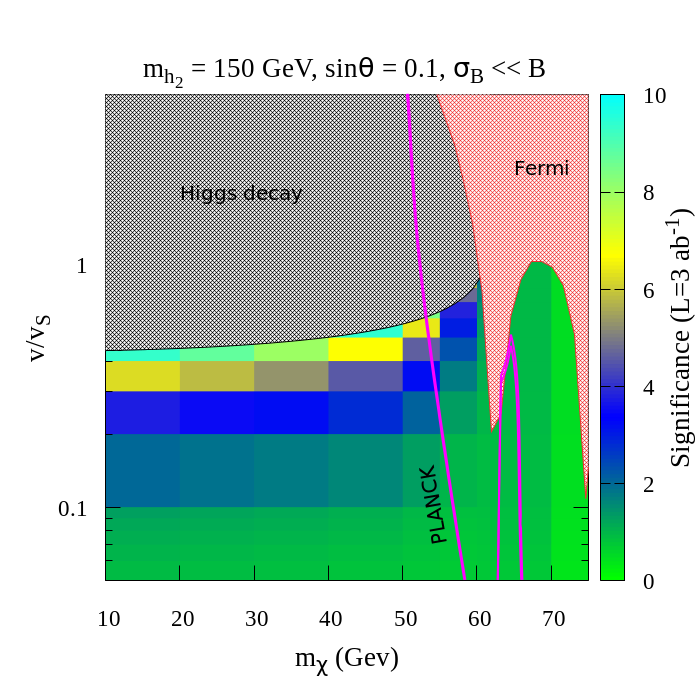}
\includegraphics[width=7cm,height=7.0cm]{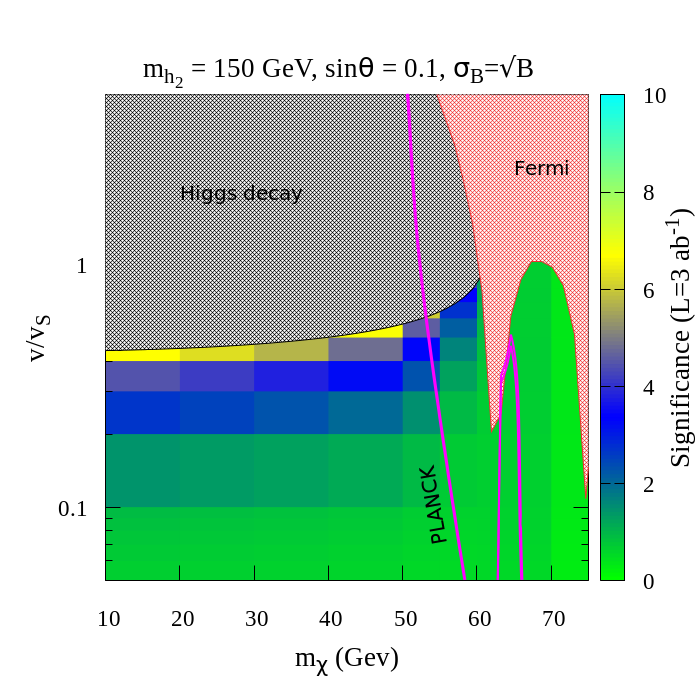}
\includegraphics[width=7cm,height=7.0cm]{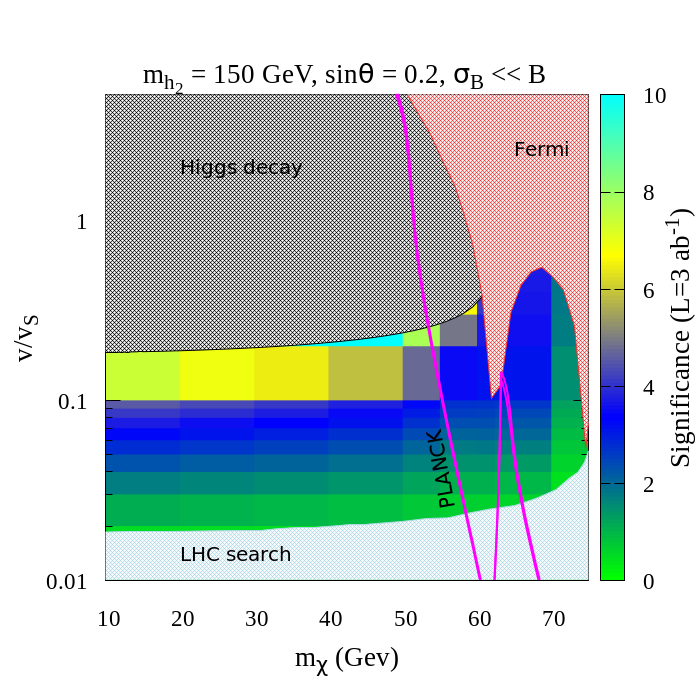}
\includegraphics[width=7cm,height=7.0cm]{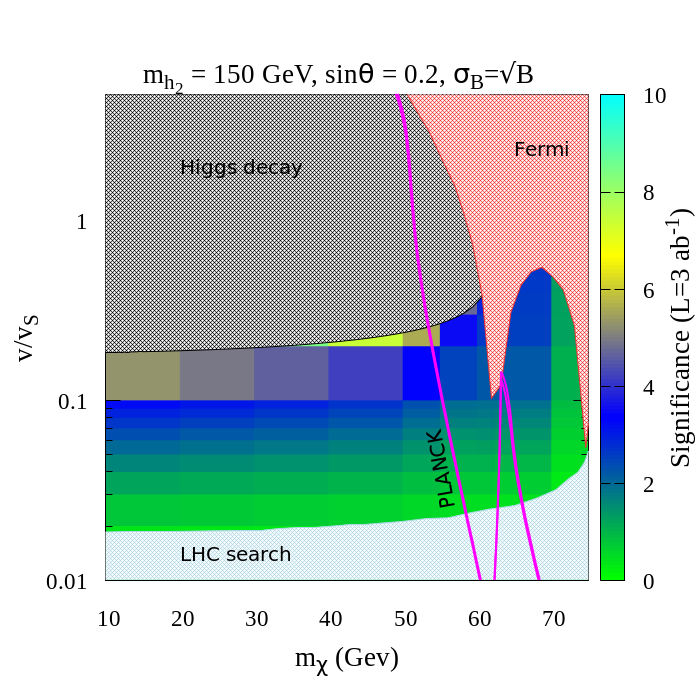}
\end{center}
\caption{Significance reach for the VBF channel with missing energy at $L=3~{\rm ab}^{-1}$
 and $m_{h_2}$=150 GeV. The shaded regions are excluded by the LHC and Fermi 
observations, while the purple band is preferred by PLANCK assuming thermal DM production. The left and right panels assume
different background uncertainties:  
$\sigma_B=0$ and $\sqrt{B}$, respectively.
\label{fig:signmh150}}
\end{figure}

In Fig.~\ref{fig:signmh150}, we show the variation of  the expected statistical significance of the VBF signal at 13 TeV 
with an integrated luminosity of $3~{\rm ab}^{-1}$ in the (${v}/{v_s}$,$m_{\chi}$) plane. The production cross-section 
of $h_1$ is much larger than that of  $h_2$, whereas its coupling to DM is much smaller. The balance between the $h_1$ 
and $h_2$ contributions to the event rate depends on the parameter choice. Let us consider a few  benchmark points (BP).
%%%%%%%%%%%%%%%%%%%%%%%%%%%%%%%%%%
%%%%%%%%%%%%%%%%%%%%%%%%%%%%%%%%%%

\begin{itemize}
\item {\bf BP1 (${\rm sin}\theta=0.2,~{v}/{v_s}=0.05,~m_{\chi}\simeq 50~{\rm GeV}$).}  The total signal cross-section at 
this point amounts to 0.26 fb with BR($h_1\to\chi\chi )\simeq 2.9\times 10^{-3}$ and BR($h_2\to\chi\chi) \simeq 0.54$.
This leads to the final state dominated by the $h_2$ decay contribution 
 accounting for $ 90\%$ of the signal events.

\item {\bf BP2 (${\rm sin}\theta=0.2,~{v}/{v_s}=0.2,~m_{\chi}\simeq 50~{\rm GeV}$).} The total signal cross-section is 
 0.93 fb with BR($h_1\to\chi\chi) \simeq 0.045$ and BR($h_2\to\chi\chi)\simeq 0.96$.  
Here we have almost equal contributions from  $h_1$ and $h_2$: 
the on-shell decay of $h_2$ accounts for $51\%$ 
of the signal events. 
Note that the smallness of the  $h_2$ production cross-section   is duly 
compensated by its large invisible decay branching ratio and slightly better cut efficiency.
\end{itemize} 

Fig.~\ref{fig:signmh150} shows that there are good prospects for light  ($m_\chi < 60$ GeV)  DM detection as long as
the background uncertainty does not substantially exceed $\sqrt{B}$. The signal significance can reach 5 and higher  for $v/v_s >0.1$.
The thermal WIMP band can only be probed  in  the vicinity of the resonance, i.e. for DM masses close to $m_{h_1}/2$.

We  find that since the off-shell $h_{1,2}$ contributions to the cross-section are small, there is no hope of probing the DM mass beyond $m_{h_2}/2$. 
%%%%%%%%%%%%%%%%%%%%%%%%%%%%%%%%%%
\begin{figure}[h!]
\begin{center}
\includegraphics[width=7cm,height=7.0cm]{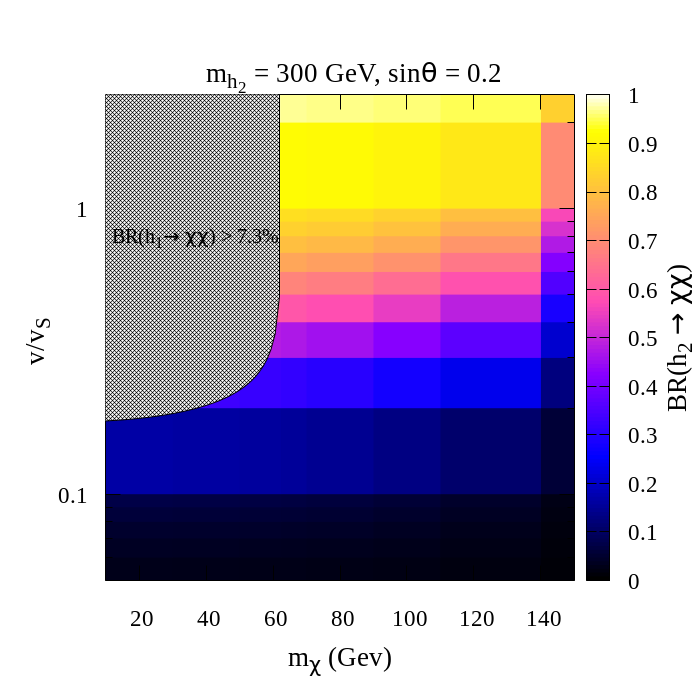}
\includegraphics[width=7cm,height=7.0cm]{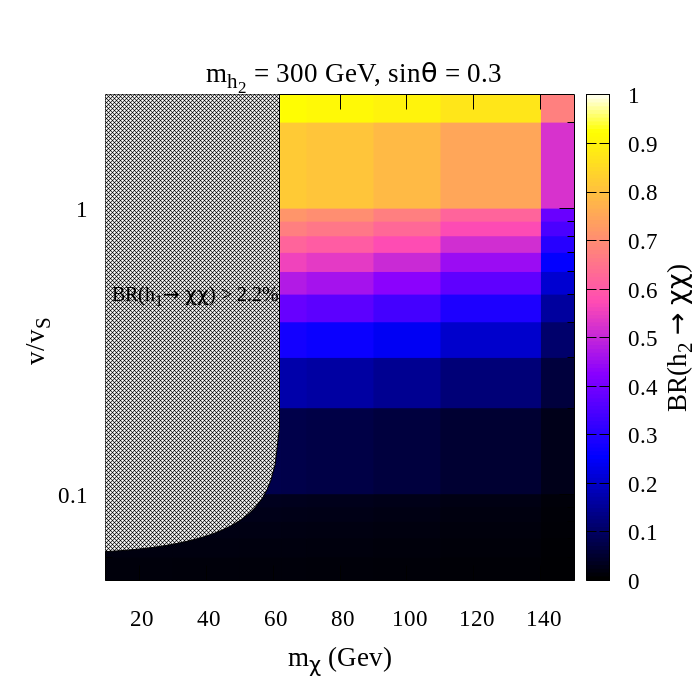}
\end{center}
\caption{BR($h_2\to\chi\chi$) and the LHC constraints for  $m_{h_2}$=300 GeV.
\label{fig:brmh300}}
\end{figure}
%%%%%%%%%%%%%%%%%%%%%%%%%%%%%%%%%%
%%%%%%%%%%%%%%%%%%%%%%%%%%%%%%%%%%
\subsubsection*{b. $m_{h_2}=300~\mathrm{GeV}$}
\label{sec:resmh300}
%%%%%%%%%%%%%%%%%%%%%%%%%%%%%%%%%%
\noindent
 Unlike in the previous case, now $h_2$ can decay on-shell into 
$WW$, $ZZ$ and $h_1h_1$ final states, with the branching ratio  for the gauge boson  modes  being the largest.
 As a result, one 
has to lower ${\rm sin}\theta$ and/or increase ${v}/{v_s}$ in order to obtain a large enough BR($h_2\to\chi\chi$). 
However, a small ${\rm sin}\theta\sim 0.1$ 
also suppresses the $h_2$ production cross-section, while a large  $v/v_s$ would be in conflict with unitarity.
Thus, we focus on  the mid-range values  $\sin\theta=0.2,0.3$.  
In Fig.~\ref{fig:brmh300}, we show the resulting  BR($h_2\to\chi\chi$) distribution.
%%%%%%%%%%%%%%%%%%%%%%%%%%%%%%%%%%
\begin{figure}[h!]
\begin{center}
\includegraphics[width=7cm,height=7.0cm]{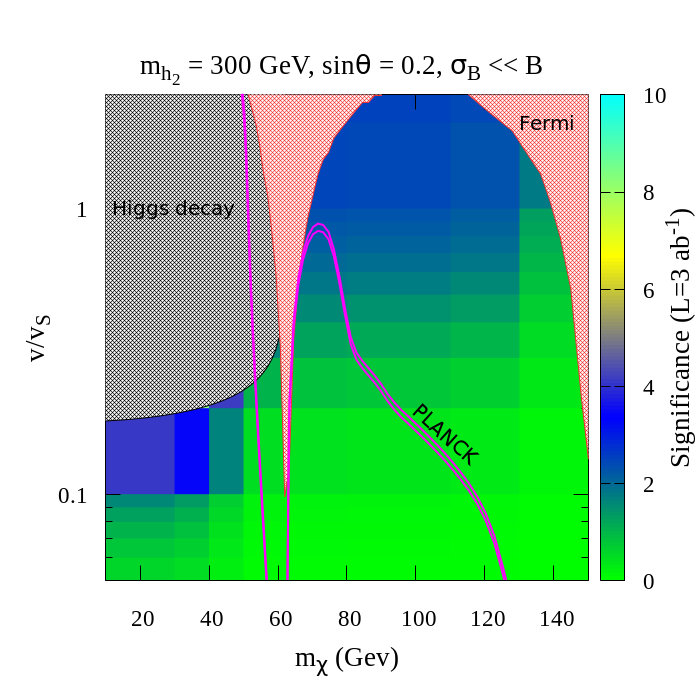}
\includegraphics[width=7cm,height=7.0cm]{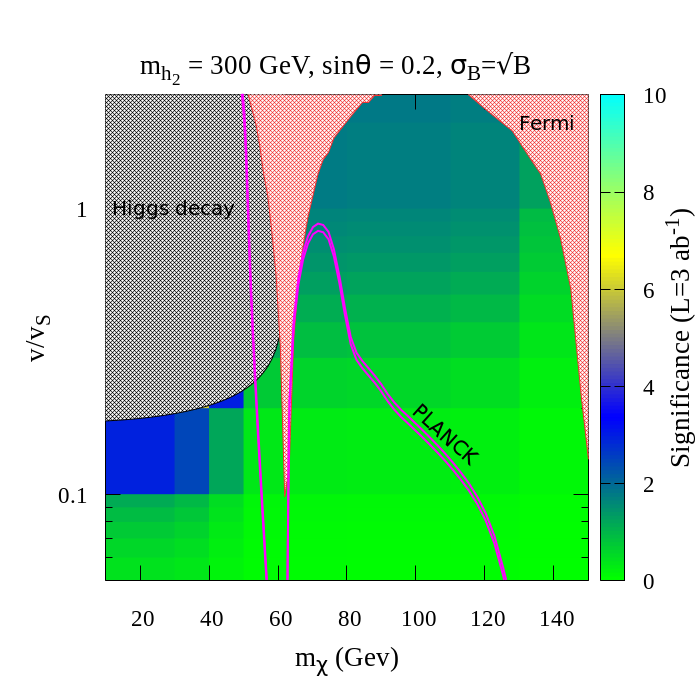}
\includegraphics[width=7cm,height=7.0cm]{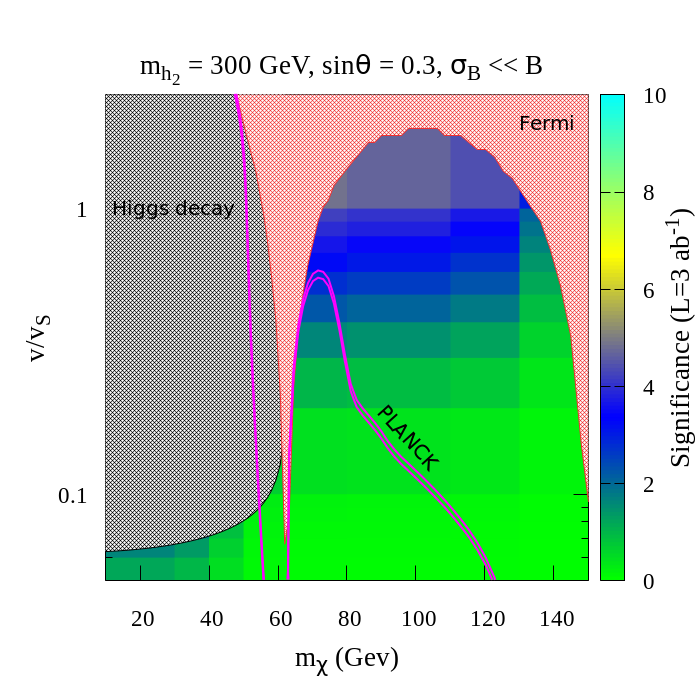}
\includegraphics[width=7cm,height=7.0cm]{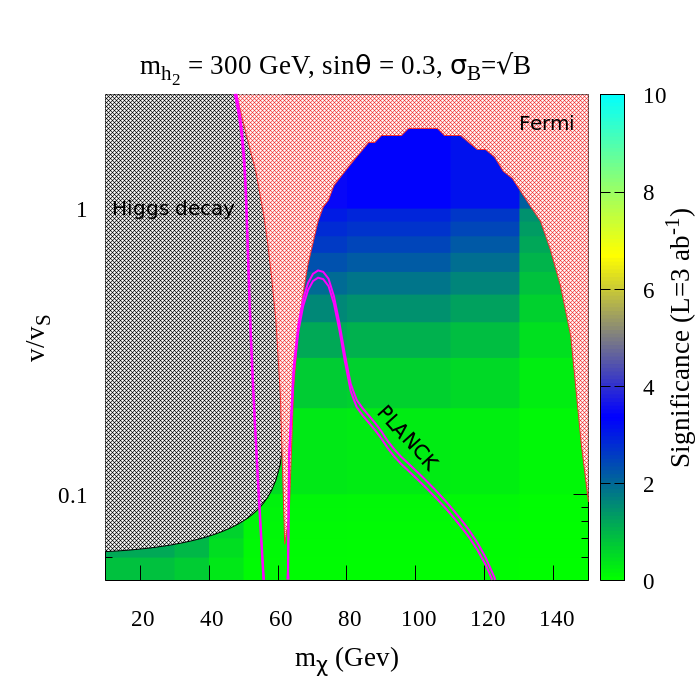}
\end{center}
\caption{ Significance reach for the VBF channel with missing energy at $L=3~{\rm ab}^{-1}$
 and $m_{h_2}$=300 GeV. The shaded regions are excluded by the LHC and Fermi 
observations, while the purple band is preferred by PLANCK assuming thermal DM production. The left and right panels assume
different background uncertainties:  
$\sigma_B=0$ and $\sqrt{B}$, respectively.
\label{fig:signmh300}}
\end{figure}
%%%%%%%%%%%%%%%%%%%%%%%%%%%%%%%%%%
 
Fig.~\ref{fig:signmh300} displays  the variation of expected statistical significance of the VBF signal at the 13 TeV
LHC with an integrated luminosity of $3~{\rm ab}^{-1}$. 
The cut efficiency improves compared to the previous case, while the cross section and BR($h_2\to\chi\chi$) reduce.
The net result is that the expected significance drops, with the maximal value  being 4 to 5 at $v/v_s \sim 1$. 

Let us consider  the benchmark points with the same $m_{\chi}$ and ${\rm sin}\theta$ as  in BP1 
and BP2 in order to assess the quantitative changes.

\begin{itemize}
\item {\bf BP3 (${\rm sin}\theta=0.2,~{v}/{v_s}=0.05,~m_{\chi}\simeq 50~{\rm GeV}$).} The total signal cross-section  
is 0.04 fb with BR($h_1\to\chi\chi) \simeq 2.9\times 10^{-3}$ and BR($h_2\to\chi\chi) \simeq 0.02$.
About $ 35\%$ of the signal events arise from the on-shell decay of $h_2$. 

\item {\bf BP4 (${\rm sin}\theta=0.2,~{v}/{v_s}=0.2,~m_{\chi}\simeq 50~{\rm GeV}$).} The total signal cross-section is 
0.57 fb with BR($h_1\to\chi\chi)\simeq 0.05$ and BR($h_2\to\chi\chi)\simeq 0.24$. 
The on-shell decay of $h_2$ accounts for $ 31\%$ of the signal events. 
\end{itemize}
Evidently, the signal rate is weaker for a heavier $h_2$. On the other hand, one can probe somewhat heavier DM up to about 120 GeV
and in certain (rather limited) regions the signal significance reaches the discovery threshold. As before, the thermal WIMP can only be probed
in the vicinity of the resonance, i.e. for $m_\chi \sim m_{h_1}/2$.

Let us close this section with a note. The CMS Collaboration has recently published its projected reach for the invisible 
decay  of the 125 GeV Higgs at high luminosity LHC. The $95\%$ CL upper bound on ${\rm BR_{inv}}$ 
assuming SM-like production of the Higgs boson is expected to be  $3.8\%$ with $3~{\rm ab}^{-1}$ integrated luminosity at 14 TeV 
\cite{CMS:2018tip}. They have optimized  the  sensitivity using the cuts $\cancel{E_T} > 190$ GeV and $|m_{jj}| > 2500$ GeV  
for this analysis. We have checked that these cuts coupled with the   ones   in Table~\ref{tab:def_sr} improve 
the significance factor in our scenario only  slightly.

%%%%%%%%%%%%%%%%%%%%%%%%%%%%%%%%%%
\section{Summary and Conclusions}
\label{sec:concl}
%%%%%%%%%%%%%%%%%%%%%%%%%%%% 
We have studied the dark matter discovery prospects in the Higgs portal framework with pseudo--Goldstone DM. 
The model is particularly attractive due to its simplicity and elegant cancellation of the direct detection 
amplitude, which allows for a wide range of DM masses consistent with XENON1T. We have focused on the VBF  
production of the Higgs--like scalars which decay into DM pairs, thereby producing the ``missing $E_T$'' signature.
Taking into account the current LHC bounds along with the indirect DM detection constraint from Fermi, we find that 
relatively light, $m_\chi \lesssim100$ GeV, dark matter can be probed in this channel with the signal significance 
at $L=\;$3 ab$^{-1}$ reaching the discovery threshold in certain regions of parameter space.

The model predicts the existence of a heavier Higgs--like boson  $h_2$ with suppressed couplings. This would provide 
a complementary test of the model, although its detection  is hindered by the strong mixing angle suppression and  a 
large invisible decay width. It is noteworthy that $h_2$ couples to dark matter much stronger than the SM--like Higgs 
$h_1$ does, hence its invisible decay can be very efficient even though the invisible decay of $h_1$  is severely constrained. 

In this scenario, dark matter can be light  quite naturally since its mass is provided by a symmetry breaking term. The direct 
detection constraints are very weak, so the lower bound  of the order of a  few  GeV is only set by the $B$--meson decays. 
Although we focus on the DM mass range above 10 GeV, essentially all of our results apply to  lower masses as well.
%%%%%%%%%%%%%%%%%%%%%%%%%%%% 
\section{Acknowledgements}
\label{sec:ack}
O.L. acknowledges support from the Academy of Finland project ``The Higgs boson and the Cosmos''.
KH, NK and SM acknowledge H2020-MSCA-RICE-2014 grant no. 645722 (NonMinimal Higgs).
T.T. is partially supported by the Grant-in-Aid for Scientific Research
on Innovative Areas from the MEXT (PI: Koji Tsumura, Grant No.18H05543).

%%%%%%%%%%%%%%%%%%%%%%%%%%%% 
\bibliography{psdm}
\end{document}